\documentclass[aps,prl,twocolumn,floats,epsfig,showpacs]{revtex4}
\usepackage{bbm}
\usepackage{amssymb}
\usepackage{ifpdf,psfrag}
\usepackage{graphicx,epsfig}

\newcommand{\eq}{\begin{equation}}
\newcommand{\en}{\end{equation}}
\newcommand{\ear}{\begin{eqnarray}}
\newcommand{\rae}{\end{eqnarray}}
\newcommand{\Z}{\mathbb{Z}}

\newcommand{\D}{{\cal D}}

\newcommand{\bra}{\langle}
\newcommand{\ket}{\rangle}

\newcommand{\tr}{{\rm tr}\,}

\begin{document}
\title{2D quantum gravity from quantum entanglement }
\author{F.\ Gliozzi}
\affiliation{
 Dipartimento di Fisica Teorica, Universit\`a di Torino, and\\
 INFN, Sezione di 
Torino, P.\ Giuria 1, 10125 Torino, Italy }

\begin{abstract}
In quantum systems with many degrees of freedom the replica method is a 
useful tool to study the entanglement of  arbitrary 
spatial regions. We apply it in a way which allows them to back-react. As a consequence, they become dynamical subsystems whose position, 
form and extension is determined by their interaction with the whole system. 
We analyze in particular quantum spin chains described at criticality by 
a conformal field theory (CFT). Its coupling to the Gibbs' ensemble of 
all possible subsystems is relevant and drives the system into a 
new fixed point which is argued to be that of the 2D quantum gravity 
coupled to this system. 
Numerical experiments on the critical Ising model show 
that the new critical exponents agree with those predicted 
by the formula of Knizhnik, Polyakov and Zamolodchikov.  
\end{abstract} 


\maketitle

Understanding  the effects of quantum entanglement of systems with many degrees of freedom such 
as quantum spin chains or quantum field theories is 
a challenging problem which connects statistical mechanics to quantum 
information science.
If the system under study is taken to be in a pure state $\vert\Psi\ket$,
 a complete description 
of the information available to an observer 
who has access only to a subsystem $A$ will be given 
by the reduced density matrix
\eq
\rho_A=\tr_B\,\vert\Psi\ket\bra\Psi\vert
\en
obtained by tracing over the degrees of freedom 
of the remainder $B$, inaccessible to the observer.

There are various functions of $\rho_A$ which could be used as useful 
probes to measure how closely 
entangled, or how `quantum', a given state is (see \cite{rev} for reviews).
Most of them are expressed in terms of $\tr \rho_A^n$. For instance, 
the R\'enyi entropy is $R_A(n)=\log\tr\rho_A^n/(1-n)$ and the entanglement entropy  is the limit $S_A=\lim_{n\to 1}R_A(n)=-\tr\rho_A\log\rho_A$.

In quantum field theory  the quantity  $\tr \rho_A^n$ for integer $n$ 
can be computed 
without explicit knowledge of the ground state through the so-called
replica method \cite{cw,hlw,cc}. Following this procedure the 
partition function $Z$ of a $d$ dimensional quantum system is computed in 
the standard way by doing the functional integration on $n$ copies 
(or replicas) of the corresponding Euclidean classical system 
in $d+1$ dimensions. 
These copies interact among themselves through the inaccessible subsystems 
$B$ as explained below. In this set up the quantity $\tr\rho^n_A$ is 
proportional to the canonical partition function $Z_n(A)$ of the coupled 
system of $n$ replicas. 
More precisely we have $\tr\rho^n_A={Z_n(A)}/{Z^n}$.     

 In most previous studies the accessible subsystem  $A$ is chosen to be fixed
 and a major goal  is to investigate how  $\tr\rho^n_A$ and the entanglement entropy depend on 
$A$. The point of view which is taken in this paper is different. We treat 
 $A$ (or equivalently $B$) as a back-reacting, dynamical, subsystem whose position, form and 
extension is determined by its interaction with the whole 
system. We implement it by `summing  over  histories', i.e. 
by putting the system in equilibrium with the Gibbs' ensemble $\{A\}$  
of all possible subsystems.

 As we will see, when the system is put on a lattice the  
sum over the ensemble $\{A\}$ is unambiguously defined. An interesting 
consequence is that the sum over $\{A\}$  defines a non-trivial, 
translation invariant, modification of the  system uniquely 
generated by the quantum 
entanglement of the set of  all the accessible subsystems. This is 
particularly interesting when the system undergoes a critical transition.
 
One important question to address is whether
the interaction with the ensemble $\{A\}$ changes the universality class 
of the critical system i.e., in renormalization group language, whether the 
coupling to $\{A\}$ is relevant when the critical system is in a pure state.  
In this paper we answer this question for the class of $1+1$ dimensional 
quantum systems
described at criticality by a relativistic, massless, field theory, i.e. a 
conformal field theory (CFT) with central charge $c$. 
It turns out that any 
fixed point with $c>0$  becomes unstable when coupled to $\{A\}$ and that the 
system flows to a fixed point with $c=0$. A series of numerical experiments on 
the  Ising model indicates that for a suitable choice of the coupling 
parameters the new fixed point is critical. The scaling dimensions of the 
primary fields turn out to be those expected in the coupling 
of CFT to two dimensional quantum gravity. 
\begin{figure}[t]
\includegraphics[angle=-90,width=0.5\textwidth]{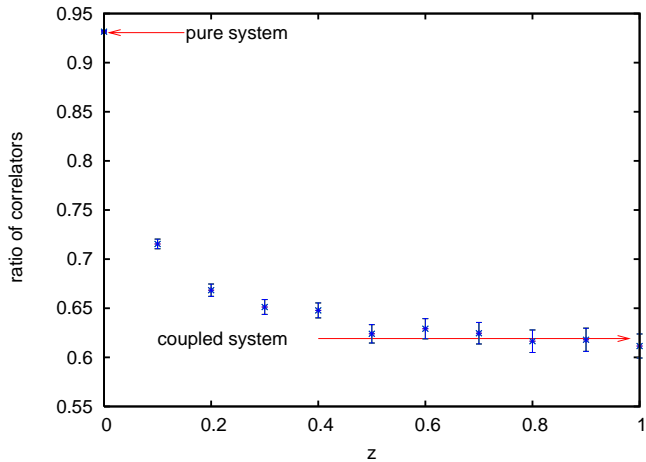}
\caption{ The  critical ratio  $\lambda^{-x}=C(\lambda L,8)/
C(L,8)$  as a function of the 
fugacity $z$ for $L=96$, $\lambda=\frac43$ and two replicas 
(see  eq.s (\ref{cls}) and (\ref{scl})). The top and the bottom arrows indicate respectively 
the expected value $\lambda^{-4\Delta^o_\sigma}$
for the pure system ($z=0$) and $\lambda^{-4\Delta_\sigma-1}$ for  the coupled 
system at $z=1$. $\Delta^o_\sigma$ and $\Delta_\sigma$ are related through 
the KPZ formula.}
\label{Figure:1}
\end{figure}

The partition function of  many   
quantum systems at inverse temperature $\beta$ may be rewritten as a 
Feynman path integral in imaginary time $0\le\tau\le\beta$.
In the case of a $d$ dimensional quantum lattice system
 we may regard the quantum partition function $Z=\tr e^{-\widehat{H}}$
 as the canonical partition function of a classical system in $d+1$ 
dimensions 
in a slab geometry with thickness $\beta$. The boundary conditions in the 
imaginary time direction are periodic for bosonic degrees of freedom.
The case when the classical system 
is infinite in all $d+1$ directions corresponds to zero temperature in the quantum system. 
For sake of simplicity we assume that the classical  system is defined on 
a $d+1$ dimensional  hyper-cubic lattice 
$\Lambda=\{\vec{x},\tau\}$  $(x_i,\tau\in
\Z)$.  
 Its partition function can be computed by doing the Euclidean functional 
integral   $Z=\int\D[\phi] e^{-S[\phi]}$    over 
fields $\phi_x\equiv\phi(\vec{x},\tau)$ periodic under $\tau\to\tau+\beta$.
The Euclidean action $S[\phi]$ is assumed to be decomposable as the sum 
$S=\sum_{\bra xy\ket}S[\phi_x,\phi_y] $ of the 
contributions of the links $\bra xy \ket$ of $\Lambda$ .

Working in the framework of replica method, we consider
a stack of $n$ copies   of the original system, range them 
in a cyclical 
order and couple them together in  the following 
way. We pick all the lattice nodes belonging to the inaccessible subsystem 
$B$ and replace their links in the imaginary time direction with links 
connecting two consecutive copies  
\eq    
S^{(k)}_{\bra xy\ket}=\cases{S[\phi_x^{(k)},\phi_y^{(k+1)}]& $x\in B$\cr
S[\phi_x^{(k)},\phi_y^{(k)}]& $x\not\in B$\cr}~~
\label{link}      
\en  
where  $\phi^{(k)}$ denotes the field in the $k-$th copy. The subsystem
$B$ lies in a slice of the lattice at a given value of $\tau$.  
Since the imaginary time and space directions enter into the problem on very different footings, 
one might expect the corresponding classical system to exhibit intrinsically anisotropic scaling. 
Here we restrict our analysis to systems which are sufficiently isotropic, in such a way that their 
critical behavior is described by a relativistic $d+1$ dimensional field theory. In this context it is 
convenient to define a slightly generalized coupling among the $n$ replicas, getting rid of the 
constraint of the $B$ subsystem to lie in a constant slice 
and   treating spatial and temporal links in the same way. Each stack of $n$ 
links associated to the nodes $x$ and $y$ in the $n$ replicas is set in two possible states. 
In the state `A' each link of the stack connects points of the same 
replica while in the state `B' 
it connects them cyclically, like in (\ref{link}); the links in the state `B' single out the 
subsystem $B$. An advantage of this more general setting is that it is easy 
to show that the coupled 
system of $n$ replicas is endowed with an important local symmetry: 
flipping from `A' to `B' or vice 
versa the state of  all links intersecting an arbitrary closed $d$ dimensional 
manifold keeps invariant the partition function \cite{cg}. 
A direct consequence of such a symmetry is that not only the entropy but all 
the thermodynamic functions depend only on the boundary of $B$ \cite{cg}.

In order to promote accessible subsystems to dynamical variables one has 
simply to sum 
 over all possible assignments of the states `A' and `B' to the lattice 
links, so the partition function of our coupled system of $n$ replicas 
can be written as
\eq
Z_n=\sum_{\{G\}}\int\prod_{k=1}^n\D\phi^{(k)}
e^{-\sum_{k=1}^n\sum_{\bra xy\ket}S^{(k)}_{\bra xy\ket}}
\label{zn}
\en 
where $G$ is the subgraph of links which are set in the state 
`B' and the summation is over all subgraphs. 

In the spirit of replica method, the evaluation of the entanglement 
entropy would require taking the limit $n\to1$, even if there are indications
that the analytic continuation from positive integer $n$ to real values 
could be rather difficult \cite{gt}. Fortunately, we do not need to do that, 
because the new phenomenon we want to describe can be observed for
any integer $n>1$. 

To make the discussion concrete and explicit, we specialize now to the case 
where the  system in question is a quantum spin chain described at criticality by a CFT. Its two dimensional lattice description 
(\ref{zn}) is a discretized version of a $n-$sheeted covering of the plane, 
where the dual $\widetilde{G}$ of the subgraph $G$  is formed by the set of cuts
connecting these sheets. The local symmetry mentioned above 
turns out to express  
the invariance of the system under the addition (or the removal) of 
closed cuts or under continuous deformations of open cuts with fixed ends. 
It is worth noting that the only elements of the graph
 $\widetilde{G}$ having an intrinsic geometrical -and physical- meaning  are the end points of the cuts, i. e. the branch points of the Riemann surface. 
They correspond  to conical singularities with deficit 
angle $2\pi(n-1)$. On the contrary, the cuts joining different branch points 
are in no way distinguished lines on the surface: their introduction has 
a similar role as the choice of a reference frame on the surface. 
Thus, the dynamical effects of the back-reaction of the 
accessible subsystems of a  critical quantum spin chain are 
intimately related to  a 2D CFT 
in statistical equilibrium with a gas of conical singularities. 

Note that the $n$-sheeted 
covering of the plane with $N$ branch points is a Riemann surface of genus 
$g=(n-1)(N-2)/2$, therefore summing over 
all accessible subsystems corresponds, in the replica approach, to a 
double sum over genera and moduli of these Riemann
surfaces. This is the first indication that this issue is related to 2D
quantum gravity.

A stronger indication comes from an observation made 
by Knizhnik long time ago \cite{k}, namely that  branch points correspond 
to primary fields $\Phi_n(z,\bar{z})$ of scaling dimensions
\eq
\Delta_n=\bar{\Delta}_n=\frac{c}{24}\left(1-\frac1{n^2}\right)~,
\label{deltan}
\en 
where $c$ is the central charge of the system at the critical point. We  
see that for $c$ not too large the conical singularities are associated to   
relevant operators. Therefore the effect of considering the accessible subsystems  
as dynamical quantities is equivalent, in the underlying field theory, 
to perturbing the CFT by the relevant operator  
$\Phi_n(z,\bar{z})$, so the action is
\eq
S=S^\star+\mu\int \Phi_n(z,\bar{z})\, d^2 z
\en  
where $S^\star$ is the CFT action, and $\mu$ is the chemical potential which 
controls the appearance of conical singularities. Such a perturbation drives 
the system away form the critical point. According to the $c-$theorem 
\cite{cth}, a generic relevant operator  generates renormalization group 
flows into fixed points with a value of $c$ that cannot exceed 
its initial value. However, there is more 
information to be gained. According to (\ref{deltan}) any CFT with $c>0$
is unstable whenever perturbed with $\Phi_n$. This  restricts the 
possible fixed points to which any CFT with $c>0$ may flow to those with 
$c=0$. For generic values of $\mu$ these are trivial fixed points corresponding to massive theories. There is however one value of $\mu$, at least, 
at which the system is critical. Thus the total central charge of the CFT
 coupled to the ensemble of the accessible subsystems is zero, which is precisely what it happens in CFT's coupled to quantum gravity. 
\begin{figure}[t]
\includegraphics[angle=-90 ,width=0.5\textwidth]{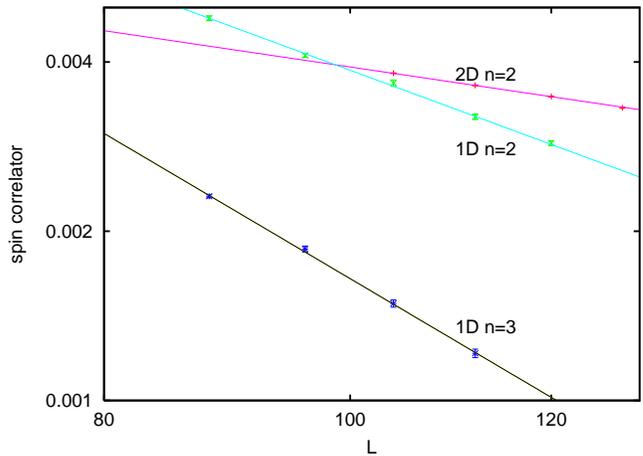}
\caption{ The two bottom lines are 
 one-parameter  fits of $(C(L,4)+C(L,8))/2$ in the 1D setting at $z=1$  
 for $n=2$ and $n=3$. The top line is a similar fit 
 in the 2D setting at $z=z_c$ for $n=2$. In this doubly logarithmic plot the 
data lie  on straight lines. The slope of the  top line is exactly 
$-4\Delta_\sigma$ while in the 1D settings is shifted by the integer 
$1-n$.  In order to represent all data in a single plot, the 2D data 
are reduced by a factor of 10. }
\label{Figure:2}
\end{figure}

The strongest indication that this new universality class corresponds to 2D quantum gravity comes from numerical experiments, where 
one may accurately evaluate how change the scaling dimensions of local 
operators when the  coupling to  accessible subsystems is switched 
on. We use as a guidance the formula of Knizhnik, Polyakov and Zamolodchikov
(KPZ)\cite{kpz,ddk} 
\eq
\Delta^o=\Delta+\frac{\gamma^2}4\Delta(\Delta-1),~~
 \gamma=\sqrt{\frac{25-c}6}-\sqrt{\frac{1-c}6} ~,
\en
 which relates the 
scaling dimensions $\Delta^o$ of a primary field of a CFT to the scaling 
dimensions $\Delta$ of the same operator when the theory  is coupled to 
quantum gravity.  

Let us consider a  spin-$\frac12$ quantum chain coupled to a 
transverse magnetic field $h$. 
The quantum Hamiltonian is
 $\widehat{H}=-J\sum_i\sigma^z_i\sigma^z_{i+1}-h\sum_i\sigma^x_i~,$
where $\sigma^x$ and $\sigma^z$ are the usual Pauli matrices.  
As is well known, this system exhibits a quantum phase transition for 
$J=h$. This manifests itself as a power law decay of the correlators 
which lies in the universality class of the 2D critical Ising 
model, described by the CFT with $c=\frac12$. 

We simulated this system at the self-dual point with 
$n=2,\dots,5$ replicas of 
a square lattice enclosed in a square box with a side of  $L$  lattice spacings 
with toroidal boundary conditions. 
In a first set of Monte Carlo calculations the ensemble $\{A\}$  was taken 
on a 1D slice  at a fixed value of $\tau$. 
On the same temporal slice we measured the (difference of) correlators
\eq
C(L,s)=\bra\sigma^z_i\sigma^z_{i+L/s}-\sigma^z_i\sigma^z_{i+L/2}\ket~,
\label{cls}
\en
where $s$ is a (proper) factor of the integer $L$. In our simulations we 
chose $s=8$ and $s=4$. In order to control the coupling with the  
ensemble $\{A\}$, we modified the Ising model by allowing a non-zero fugacity 
$z=e^{-\mu}$ counting the number of branch points.  

\begin{figure}[t]
\includegraphics[angle=-90 ,width=0.5\textwidth]{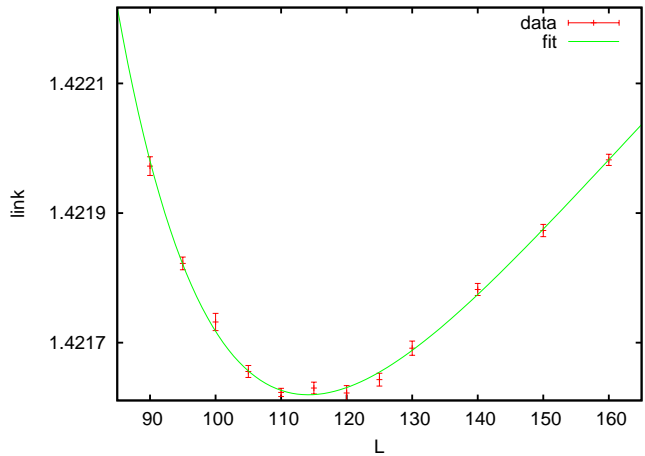}
\caption{ The link variable as a function of the size of the 
lattice. The solid curve is a fit to eq.(\ref{energy}).}
\label{Figure:3}
\end{figure}
Since $\{A\}$ affected only a one dimensional boundary of our system, 
this remained critical even at $z\not=0$. We simulated the Ising part of the 
model with a standard non-local cluster algorithm, while the update of the 
subset $A$, consisting of an arbitrary number  $N$ of disjoint intervals 
with $N=0,1,\dots ,L/2$, was performed with a heat bath method.
Although the specific form of $C(L,s)$ is not known, we have, 
in the thermodynamic limit,
\eq
C(\lambda L,s)=\lambda^{-x} C(L,s)
\label{scl}
\en   
where $\lambda$ is any positive rescaling factor. As $z$ varies from 0 to 1, 
$x$ drops from the expected value of the pure Ising model 
$x=4\Delta^o_\sigma$ to a new scaling dimension (see figure 1). According to 
the previous discussion, the expected value is the one suggested by KPZ 
formula.  
As sometimes it happens in critical systems, the observed value of $x$ 
differs by an integer with respect the expected value. Precisely
we found $x=4\Delta_\sigma+1$ for $n=2$  and  
$x=4\Delta_\sigma+2$ for $n\ge 3$ (see figures 1 and 2).  
$\Delta_\sigma=\frac16$ is the KPZ value associated to 
$\Delta_\sigma^o=\frac1{16}$ for $c=\frac12$. 

Let us emphasize that in the first set of numerical experiments 
the simulated system is one dimensional, in the sense that the correlator and 
the subsystems are taken on a fixed 1D slice at a given value of $\tau$. 
In the second set of numerical experiments we simulated instead a truly 
two dimensional system, with no limitations on the location of cuts 
representing the accessible subsystems. In this new setting  the gas 
of conical singularities is spread in the bulk  and drives the system 
away from the critical point of the pure system. There is however 
a critical value $z_c$ of the fugacity at which  the whole interacting  
system undergoes a second order phase transition which is presumably 
in the same universality class of the 1D quantum system described above. 
The system composed of two replicas 
at the self-dual point of the pure Ising system exhibits a critical behavior 
for $z_c=0.01127(1)$. The scaling dimension of the spin operator   turns out to be, within the numerical accuracy, the one of the 2D quantum 
gravity (see figure 2). Likewise, the scaling dimensions of 
the energy operator turn 
out to be the KPZ value $\Delta_\epsilon =\frac23$, corresponding 
to $\Delta^o_\epsilon=\frac12$. In order to extract this critical exponent, 
we measured the vacuum expectation value of the link operator 
$\bra\sigma^z_x\,\sigma^z_y\ket$, where $x$ and $y$ represent
two nearest neighbors of the lattice. At the critical point this 
quantity is expected to have the following functional form
\eq
 \bra\sigma^z_x\,\sigma^z_y\ket=e_0+e_1/L^{2\Delta_\epsilon}+
e_2/L^{2\Delta_\epsilon+1}+ \dots
\label{energy}
\en
where we wrote explicitly only the terms necessary to  accurately fit 
the data, as shown in figure 3.

From a geometrical point of view, the two-sheeted system we 
described is defined on an hyperelliptic Riemann surface and the 
sum $\sum_{\{G\}}$ over all possible subsets of links  made in eq.(\ref{zn}) 
corresponds, in the continuum limit, to a double sum over the space of  moduli 
and the genera of these surfaces. This has some similarity with the double 
scaling limit of matrix models \cite{df}. A promising aspect of the present 
approach is that it may be extended in a straightforward way to higher 
dimensions.
 
Let us conclude with a general remark. In the past years, 't Hooft 
\cite{'t Hooft:1984re} and several other authors \cite{bom,sr,ks,cw} have 
suggested that the quantum entanglement and its ensuing entropy might be related to the Bekenstein-Hawking entropy of black holes. More recently, in the 
light of the AdS/CFT correspondence, a comprehensive gravitational 
interpretation of the entanglement entropy has been 
proposed \cite{Ryu:2006bv}. The present study captures a different facet of the same fascinating relationship  between gravity and  quantum entanglement, in 
which the back-reaction of the accessible subsystems of a 1D quantum system 
has the same effect as the coupling to 2D quantum gravity.

\end{document}